\documentclass[letterpaper]{jps-cp}
\usepackage{txfonts} 

\newcommand{\mSR}{$\mu$SR}

\newcommand{\scg}{superconducting}

\newcommand{\CBS}{Cu$_x$Bi$_2$Se$_3$}
\newcommand{\BS}{Bi$_2$Se$_3$}
\newcommand{\lam}{$\lambda_{\mathrm{eff}}$}

\usepackage{multirow}

\title{Superconducting properties of Cu intercalated Bi$_2$Se$_3$ studied 
by Muon Spin Spectroscopy}

\author{
Jonas~A.~\textsc{Krieger}$^{1,2}$,
Amit~\textsc{Kanigel}$^{3}$,
Amit~\textsc{Riback}$^{3}$,
Ekaterina~\textsc{Pomjakushina}$^{4}$,
Khanan~B.~\textsc{Chashka}$^{3}$,
Kazimierz~\textsc{Conder}$^{4}$,
Elvezio~\textsc{Morenzoni}$^{1}$,
Thomas~\textsc{Prokscha}$^{1}$,
Andreas~\textsc{Suter}$^{1}$,
and 
Zaher~\textsc{Salman}$^{1*}$
}

\inst{
$^{1}$Laboratory for Muon Spin Spectroscopy, Paul Scherrer
Institute, CH-5232 Villigen PSI, Switzerland \\
$^{2}$Laboratorium f\"ur Festk\"orperphysik, ETH-H\"onggerberg, CH-8093
Z\"urich, Switzerland\\
$^{3}$Department of Physics, Technion - Israel Institute of Technology, 
Haifa 32000, Israel\\
$^{4}$Laboratory for Scientific Developments and Novel Materials, Paul Scherrer Institute, 
CH-5232 Villigen PSI, Switzerland\\
}

\email{zaher.salman@psi.ch}

\abst{ We present muon spin rotation measurements on \scg\ Cu
  intercalated Bi$_2$Se$_3$, which was suggested as a realization of a
  topological superconductor. We observe a clear evidence of the \scg\
  transition below 4 K, where the width of magnetic field distribution
  increases as the temperature is decreased. The measured broadening
  at mK temperatures suggests a large London penetration depth in the
  $ab$ plane (\lam$\sim 1.6$~$\mathrm{\mu}$m). We show that the
  temperature dependence of this broadening follows the BCS
  prediction, but could be consistent with several gap symmetries.}

\kword{low energy muons spin spectroscopy, topological
superconductivity, \CBS}

\begin{document}
\maketitle

Similar to topological insulators which have an insulating bulk and
topologically protected conducting surface states, a topological
superconductor features a bulk \scg\ gap with gapless boundary
states~\cite{Sato2017}. The quasiparticle excitations in these surface
states can act as Majorana fermions, which are particles that are
their own antiparticles~\cite{Sato2016}. In addition, it has been
predicted that spin triplet $p$-wave superconductivity can be produced
at the interfaces between an $s$-wave superconductor and a topological
surface
state~\cite{Fu2008,Stanescu2010,Potter2011,BlackSchaffer2013}. A \scg\
vortex at such an interface is expected to contain a Majorana bound
state, that is the basic element in a recent proposal for
fault-tolerant quantum computing \cite{Kitaev2006}. The prototypical
topological insulator Bi$_2$Se$_3$ has been reported to become \scg\
upon carrier doping by either Sr, Nb or Cu
intercalation~\cite{Liu2015c,Smylie2016,Hor2010}.  It has been shown
with angle resolved photoemission spectroscopy (ARPES) and torque
magnetometry that the Dirac surface states of the parent material are
inherited by \CBS~\cite{Wray2010,Lawson2012,Lahoud2013}. This and the
strong spin orbit coupling in Bi$_2$Se$_3$ have made this material a
promising candidate for a topological superconductor. Theoretically,
it is proposed that odd-parity pairing, a full gap and an odd number
of time reversal symmetry (TRS) invariant momenta enclosed by the
Fermi surface are sufficient to make the superconductivity in \CBS\
topologically nontrivial~\cite{Fu2010}.  It is worth noting here that
the strong spin orbit coupling prevents the usual characterization of
the superconducting state with the symmetry of the gap function into
$s$-, $p$-, $d$-, ...  wave~\cite{Sato2017}.

Despite much work that has been dedicated to optimize the preparation
procedure of \CBS\ the observed diamagnetic shielding fraction remains
below 70\%, which indicates partial volume
superconductivity~\cite{Kriener2011b,Schneeloch2015,Wang2016}.  First
indications for topological superconductivity in \CBS\ were reported
by point contact spectroscopy, where a zero bias conductance peak
(ZBCP) indicated the presence of Majorana fermions at the
surface~\cite{Sasaki2011,Kirzhner2012,Ando2013}. In contrast, no ZBCP
was found with scanning tunnel
microscopy~\cite{Levy2013}. Furthermore, ARPES measurements and the
observation of quantum oscillation revealed that the Fermi surface
becomes two-dimensional-like with increasing carrier concentration,
thereby enclosing two TRS invariant
momenta~\cite{Lahoud2013,Lawson2014}.  Recently, new evidence of
topologically nontrivial, nematic superconductivity has been found
using NMR measurements~\cite{Matano2016}.  The $^{77}$Se Knight shift
in the superconducting state exhibits a two fold rotation symmetry
when the external field is rotated in the
$ab$-plane~\cite{Matano2016}. Similar features were also observed for
the specific heat and the upper critical
field~\cite{Yonezawa2017}. Since the rotation symmetry of the crystal
in the $ab$-plane is hexagonal, this is a strong indication that the
superconductivity is topologically nontrivial.

In this paper we present results from muon spin rotation ($\mu$SR)
measurements on Cu intercalated Bi$_2$Se$_3$ in the vortex state,
where the magnetic field is applied along the $c$-axis of the
crystal. From these measurements, we estimate the effective London
penetration depth in the $ab$-plane at mK temperatures to be $\sim
1.6$~$\mathrm{\mu}$m. Our measurements reveal that the \scg\ gap in
the bulk exhibits BCS temperature dependence, which could be
consistent with various pairing symmetries.

The \mSR\ measurements were performed on the High-field spectrometer
(HAL-9500) at the Paul Scherrer Institute, Switzerland. In these
measurements, fully spin polarized positive muons ($\mu^+$) are
implanted into the studied sample. Muons decay into a positron with a
lifetime of $\tau_\mu \sim 2.2$~$\mu$s, which is emitted
preferentially in the direction of the muon's spin at the time of
decay. The ensemble average of the polarization of the implanted muons
is proportional to the asymmetry, $A(t)$, which is monitored via the
asymmetric decay using appropriately positioned detectors. $A(t)$ can
be used to extract information regarding the internal magnetic fields
at the muon site.

The measurements reported here were performed on high quality single
crystals of \CBS. These samples were prepared using electrochemical Cu
intercalation in single crystals of \BS. The samples are similar to
those used in Ref.~\cite{Ribak2016} and have a \scg\ volume fraction
of $40-60$\% as estimated from magnetization measurements.  The
crystals were glued (using Apiezon grease) onto an 8~mm diameter
silver rod, which serves as a cold finger of a dilution fridge. The
magnetic field (applied along the $c$-axis of the \CBS\ crystals) is
generated using a \scg\ warm bore magnet (0-9.5 Tesla).

\begin{figure}[htb]
\center
\includegraphics[width=0.5\linewidth]{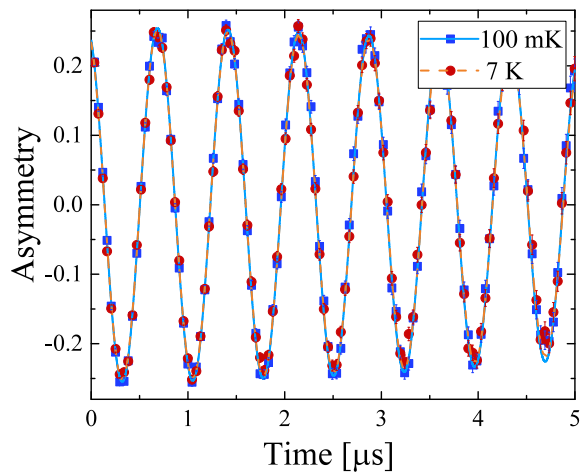} 
\caption{Typical spectra in one of the detectors. There is no visible
  difference in the spectra at temperatures above and below the
  superconducting transition. The lines show fits with a Gaussian
  damped oscillation.}
\label{fig:Asy}
\end{figure}
\begin{figure}[htb]
  \centerline{\includegraphics[width=0.5\linewidth]{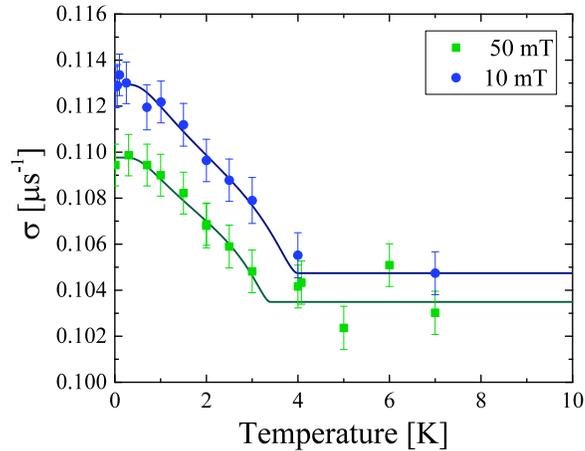}}
  \caption{The damping rate of the precessing asymmetry measured in
    Cu$_{x}$Bi$_2$Se$_3$ as a function of temperature at 10 mT
    (circles) and 50 mT (squares). The solid lines show a fit of an
    anisotropic gap with a BCS temperature dependence.}
  \label{fig:SigvsT}
\end{figure}

In the normal state of \CBS\ the internal magnetic field sensed by the
implanted muons is primarily the applied magnetic field with the
addition of a small contribution from to the magnetic moments of the
nuclei in the system. Below $T_c$, a vortex lattice is formed in SC
\CBS\ which results in an inhomogeneous broadening of the field
distribution sensed by the muons. Typical asymmetry spectra are shown
in Fig.~\ref{fig:Asy}. The difference between measurements above and
below $T_c$ is not visible by eye. To allow for an accurate fitting of
this data a total of more than $100$~million events have been
collected at each temperature.  The spectra were fitted to a Gaussian
damped oscillation using the \texttt{Musrfit} software
package\cite{Suter2012}. The damping rate, $\sigma$, which reflects
the width of the internal field distribution, as a function of
temperature is shown in Fig.~\ref{fig:SigvsT}. We detect a clear
increase in the broadening below $T_c$, with $\sigma$ saturating at
very low temperatures. This is an indication that the \scg\ state in
\CBS\ is fully gapped. The change in $\sigma$ is very small and on the
limit of what is detectable with \mSR. The broadening is expected to
slightly decrease with increasing field, as the spacing between the
vortices decreases~\cite{Brandt2003}. This is in agreement to what we
observe for measurements in different fields in Fig.~\ref{fig:SigvsT}.

In the limit of a small applied field, $\sigma$ is related to the
London penetration depth in a superconductor, such
that~\cite{Brandt1988}
\begin{equation}\label{eq:lambda}
  \sigma_{\mathrm{sc}}(T)=\sqrt{\sigma(T)^2-\sigma_{\mathrm{ns}}^2}
  = 0.0609\gamma_\mu\Phi_0/\lambda_{\mathrm{eff}}^{2}(T),
\end{equation}
where $\gamma_{\mathrm \mu} = 2 \pi \times 135.5$~MHz/T is the
gyromagnetic ratio of the muon, $\Phi_0$ is the magnetic flux quantum,
$\sigma_{\mathrm{ns}}$ is the broadening in the normal state
determined above $T_c$ and \lam\ is the effective London penetration
depth under the assumption that the whole sample is
superconducting. The temperature dependence of \lam\ can be modelled
within BCS theory~\cite{Carrington2003}. The results of fits of this
temperature dependence for various gap symmetries are summarized in
Table~\ref{tab:Results}. Due to the large statistical error on the
damping rate, with regard to the change at the superconducting
transition, all listed gap symmetries are consistent with the observed
signal.  The inferred size of the superconducting gap is in agreement
with the 0.6~meV reported by other
techniques~\cite{Wray2010,Kriener2011,Sasaki2011,Kirzhner2012}. The
penetration depth is found to be \lam$\approx1.6$~$\mu$m (for the
measurements in $10$~mT, which is closer to the small field limit
assumed in Eq.~\ref{eq:lambda}). Note that this value will be reduced
if the assumption of a fully \scg\ sample is removed. The best fit
results in terms of the reduced $\chi_{\mathrm{red}}^2$ are obtained
for a full, anisotropic gap. This is consistent with the recently
proposed fully gapped $\Delta_{4y}$ case, where the \scg\ gap has
minima along the $\mathbf{k}_x$ direction~\cite{Yonezawa2017}.
\begin{table}[htb]
    \centering
    \begin{tabular}{l|c|c|c|c|}
     Model &applied Field [mT]& Gap [meV]& \lam\ [$\mu$m] &
$\chi^2_{\mathrm{red}}$\\ \hline\hline
 \multirow{2}{*}{isotropic ($s$-wave) gap} & 10& 0.64(9) & 1.61(3) &
\multirow{2}{*}{0.208}\\
 & 50 & 0.5(1) & 1.74(6) &  \\\hline
 \multirow{2}{*}{anisotropic gap} & 10 & 0.9(3) &  1.59(3)  &
\multirow{2}{*}{0.052}\\
  & 50 & 0.8(5) & 1.71(4) & \\\hline
 \multirow{2}{*}{nodal ($d$-wave) gap} & 10 & 1.2(3) & 1.58(3) &
\multirow{2}{*}{0.070}\\
 & 50 & 1.0(5) & 1.68(7)& \\\hline\hline
     \end{tabular}
     \caption{Results of fits to different gap
       symmetries using a BCS temperature dependence. The
       small $\chi^2_{\mathrm{red}}$ show that all models are
       consistent with the observed increase in $\sigma$.}\label{tab:Results}
\end{table}

In conclusion, we have characterized the superconducting state of
\CBS\ with \mSR. We find a very long effective penetration depth in
the $ab$-plane, \lam$\approx1.6$~$\mu$m, which is on the limit of what
can be detected by \mSR. A fit to a full and anisotropic gap to the
temperature dependence of the superconducting state gives the lowest
$\chi_{\mathrm{red}}^2$. However, an isotropic or nodal gap cannot be
ruled out.

\section*{Acknowledgements}
We are grateful to Robert Scheuermann for his continuous development
of the HAL-9500 spectrometer and for performing the measurements
reported here. We would also like to thank Zurab Guguchia, Stephen
Weyeneth and Marisa Medarde for their help with the dc magnetization
measurements and Rustem Khasanov for helpful discussions. The \mSR\
measurements were performed at the Swiss Muon Source (S$\mu$S), at the
Paul Scherrer Institute in Villigen, Switzerland. The work at PSI was
partially supported by the Swiss National Science Foundation
(SNF-Grant No.~200021\_165910).

\end{document}